\documentclass[11pt]{article}  
\usepackage{amsmath}
\usepackage{amsfonts,amssymb}
\usepackage[dvips]{graphics}

\begin{document}

\newcommand{\nl}{\nonumber\\}
\newcommand{\nnl}{\nl[6mm]}
\newcommand{\nle}{\nl[-2.0mm]\\[-2.0mm]}
\newcommand{\nlb}[1]{\nl[-2.0mm]\label{#1}\\[-2.0mm]}
\newcommand{\bl}{&&\quad}
\newcommand{\ab}{\allowbreak}

\renewcommand{\leq}{\leqslant}
\renewcommand{\geq}{\geqslant}

\renewcommand{\theequation}{\thesection.\arabic{equation}}
\let\ssection=\section
\renewcommand{\section}{\setcounter{equation}{0}\ssection}

\newcommand{\be}{\bes}
\newcommand{\ee}{\ees}
\newcommand{\bes}{\begin{eqnarray}}
\newcommand{\ees}{\end{eqnarray}}
\newcommand{\eens}{\nonumber\end{eqnarray}}
\newcommand{\barr}{\begin{array}}
\newcommand{\earr}{\end{array}}

\renewcommand{\/}{\over}
\renewcommand{\d}{\partial}
\newcommand{\dst}{d^2\!\si\,d\tau\ }

\newcommand{\vect}{{\mathfrak{vect}}}
\newcommand{\map}{{\mathfrak{map}}}
\newcommand{\mapp}{\map(n,p,\oj)}
\newcommand{\vmap}{\vect(n)\ltimes \mapp}

\newcommand{\eps}{\epsilon}
\newcommand{\dlt}{\delta}
\newcommand{\al}{\alpha}
\newcommand{\bt}{\beta}
\newcommand{\gm}{\gamma}
\newcommand{\z}{\zeta}
\newcommand{\si}{\sigma}
\newcommand{\vsi}{\varsigma}
\newcommand{\Si}{\Sigma}
\newcommand{\la}{\lambda}
\newcommand{\La}{\Lambda}
\newcommand{\ka}{\kappa}

\newcommand{\xmu}{\xi^\mu}
\newcommand{\xnu}{\xi^\nu}
\newcommand{\ynu}{\eta^\nu}

\newcommand{\dmu}{\d_\mu}
\newcommand{\dnu}{\d_\nu}
\newcommand{\drho}{\d_\rho}
\newcommand{\dsi}{\d_\si}
\newcommand{\dtau}{\d_\tau}

\newcommand{\dxmdt}{{\d x^\mu\/\d\tau}}

\newcommand{\mn}{{\mu\nu}}
\newcommand{\mnr}{{\mu\nu\rho}}
\newcommand{\mr}{{\mu\rho}}
\newcommand{\ms}{{\mu\si}}
\newcommand{\mt}{{\mu\tau}}
\newcommand{\nr}{{\nu\rho}}
\newcommand{\rs}{{\rho\si}}
\newcommand{\st}{{\si\tau}}
\newcommand{\sm}{{\si\mu}}

\newcommand{\ethmn}{\eth_\mn}
\newcommand{\ethmr}{\eth_\mr}
\newcommand{\ethnr}{\eth_\nr}
\newcommand{\ethrs}{\eth_\rs}
\newcommand{\ethst}{\eth_\st}

\newcommand{\hmu}{\hat\mu}
\newcommand{\hnu}{\hat\nu}
\newcommand{\one}{\hat1}
\newcommand{\two}{\hat2}
\newcommand{\three}{\hat3}

\renewcommand{\L}{{\cal L}}
\newcommand{\J}{{\cal J}}
\renewcommand{\P}{{\cal P}}
\newcommand{\R}{{\cal R}}
\newcommand{\Lxi}{\L_\xi}
\newcommand{\JX}{\J_X}
\newcommand{\RF}{\R_F}

\newcommand{\tT}{\tilde T}
\newcommand{\tU}{\tilde U}
\newcommand{\tLa}{\tilde \La}

\newcommand{\End}{{\rm End}}
\newcommand{\id}{{\rm id}\kern0.1mm}
\newcommand{\tr}{{\rm tr}\kern0.7mm}
\newcommand{\oj}{{\mathfrak g}}
\newcommand{\der}{{\mathfrak{der}}}
                                            
\newcommand{\no}[1]{{\,:\kern-0.7mm #1\kern-1.2mm:\,}}

\newcommand{\mm}{{\mathbf m}}
\newcommand{\rr}{{\mathbf r}}

\newcommand{\RR}{{\mathbb R}}
\newcommand{\CC}{{\mathbb C}}
\newcommand{\ZZ}{{\mathbb Z}}
\newcommand{\NN}{{\mathbb N}}

\title{{ Gerbes, covariant derivatives, $p$-form lattice gauge theory, 
 and the Yang-Baxter equation}}

\author{T. A. Larsson \\
Vanadisv\"agen 29, S-113 23 Stockholm, Sweden\\
email: thomas.larsson@hdd.se}

\maketitle 
\begin{abstract} 
In $p$-form lattice gauge theory, the fluctuating variables live on
$p$-dimensional cells and interact around $(p+1)$-dimensional cells. It has
been argued that the continuum version of this model should be described
by $(p-1)$-gerbes. However, only connections and curvatures for gerbes are
understood, not covariant derivatives. Using the lattice analogy, an
alternative definition of gerbes is proposed: sections are functions
$\phi(x,s)$, were $x$ is the base point and $s$ is the surface element. In
this purely local formalism, there is a natural covariant derivative. The
Yang-Baxter equation, and more generally the simplex equations, arise as
zero-curvature conditions. The action of algebras of vector fields and
gerbe gauge transformations, and their abelian extensions, are described.
\end{abstract}

\renewcommand{\arraystretch}{1.4}

\section{Introduction}

Gerbes have recently attracted considerable attention. Apart from being
of intrinsic mathematical interest \cite{BM01,Bry93,Hit99}, they appear
in Hamiltonian quantization \cite{CMM97}
and in brane models \cite{CMR02,Zun00}.
There also seem to be close ties between gerbes and n-category theory
\cite{BD95,Fre94}

My interest in this subject arose when I tried to understand lattice 
integrable statistical models. As is well known, the quantum Yang-Baxter 
equation (QYBE) is a sufficient (and in practise necessary) condition 
integrability in two dimensions \cite{Bax82}. The analogous sufficient 
condition in three dimensions is Zamolodchikov's tetrahedron equation
\cite{BS84,Bax83,BF86,Zam81}, and in $n$ dimensions it is the so-called
$n$-simplex equation. Unfortunately, very few solutions to the 
tetrahedron equation are known, and those that are known do not depend
on any temperature-like parameters.

Long ago I observed a striking similarity between
the QYBE and the zero-curvature condition in lattice gauge theory. This
lead me to formulate a statistical model where the fluctuating variables
live on plaquettes rather than links \cite{Lar90}. In section 2 this
model is reviewed, and a serious flaw is corrected: it is necessary to
assign several variables to each plaquette. An important feature of this
model is that the gauge-invariant holonomy is given by two-dimensional
``Wilson surfaces''. The generalization to higher-dimensional models 
is now obvious: the variables live on $p$-cells and there is holonomy 
associated to $p$-dimensional ``Wilson submanifolds''.

The continuum limit of this kind of model is $p$-form electromagnetism in
the abelian case, and in general it is a gauge theory 
on loop space. The first such model was probably written down by Freund
and Nepomechie \cite{FN82}, but already a few years before had Polyakov
noted that ordinary gauge theory can be formulated as a sigma model in
loop space \cite{Pol79,Pol87}. This immediately suggests that continuum
integrability in higher dimensions is best understood in terms of flat
connections on gerbes, which is closely related to BF theory \cite{AFSG97}.

Whereas a theory of connections on abelian gerbes has been around for a
few years \cite{Hit99,MacPi01}, 
the situation for non-abelian gerbes is very tentative. 
Various suggestions, whose mutual relations are unclear to me, can be
found in \cite{Att01,Att02,BM01,Mac01}.
What does not seem right, neither from the loop space approach or from
lattice gauge theory, is the appearence of a hierarchy of connections.
Another problem with the gerbe and BF approaches is that there seems to
be no natural covariant derivative. This is a serious drawback since the
main interest in connections is their relation to parallel transport.
The standard approach to gerbes is reviewed in Section 3.

To remedy these problems, I suggest an alternative definition of gerbes
in Section 4. The main idea is that a gerbe section $\phi(x,s)$ is a
function not only of the spacetime point $x\in\RR^n$, but also
a function of the direction $s$. A gerbe section is a functional on
loops over $\RR^n$, the local information about which is encoded in 
the point $x$ and the direction $s$ of the loop passing through $x$.
The generalization to higher gerbes is obvious, and in fact we mainly 
formulate the results for $2$-gerbes. In this way, a manifestly local
tensor calculus is developed, which should be useful for explicit
calculations. We construct covariant derivatives, connections and
curvatures and check that all constructions are invariant both under
infinitesimal spacetime diffeomorphisms and under gerbe gauge 
transformations.

Flatness of the gerbe connection gives rise to the classical Yang-Baxter 
equation in a particular gauge, which again indicates relevance for 
higher-dimensional integrability. Finally, it is shown how $p$-form
electromagnetism can be recovered in the abelian case. The language
developed in the present paper is thus appropriate for the non-abelian
generalization of $p$-form electromagnetism, which is known to be 
impossible using ordinary bundles \cite{Tei86}.

I thank Peter Orland for pointing out early references
\cite{Nep83,Orl83,Orl84a,Orl84b}, where lattice models very similar to
that in Section 2 were constructed.

\section{A generalization of lattice gauge theory}

Recall that in ordinary lattice gauge theory \cite{Kog79}, 
the amplitude for parallel transport along a link from $x$ to $x+\hmu$ 
is the matrix
\be
U_\mu(x) = \exp(i a A_\mu(x)),
\ee
where $a$ is the lattice spacing and $A_\mu$ is the $\mu$:th component of 
the gauge potential. Here $x\in\ZZ^n$ is a point on an $n$-dimensional
hypercubic lattice and $\hmu$ denotes a unit vector in the $\mu$:th 
direction. If a particle is transported along a link and then
back again, nothing has happened, so we associate the matrix 
$U_{-\mu}(x+\hmu) = U_\mu^{-1}(x)$
with the same link with the opposite orientation. The curvature is given 
by the holonomy around a plaquette, which is the smallest loop that can be 
constructed.
\be
U_\mu(x) U_\nu(x+\hmu) U_\mu^{-1}(x+\hnu) U_\nu^{-1}(x)
= \exp(i a^2 F_\mn(x)).
\ee
The action reads 
\be
\sum_{\hbox{plaquettes}} \tr UUUU + h.c. 
= const + a^2 \int d^nx\, F_\mn(x)F^\mn(x) + O(a^3).
\label{SYM}
\ee
We see that in the limit that the lattice spacing $a \to 0$, this
becomes the Yang-Mills action, at least formally.
There is a gauge symmetry associated to each vertex; the transformation
\be
U_\mu(x) \to f^{-1}(x+\hmu)U_\mu(x) f(x)
\ee
leaves the action (\ref{SYM}) invariant. 
The gauge invariant observables are Wilson loops, i.e. the product of 
matrices around a closed loop.

Instead of putting matrices on links, it is natural to put four-index
quantites on plaquettes. Associate a vector space $V$ to each link,
and an element $U_\mn \in \End(V_\mu \otimes V_\nu)$ to a plaquette in the 
$\mn$-plane. $U_\mn$ is in fact an element acting on 
$V \otimes V \otimes ... \otimes V = V^{ \otimes n}$, where $n$ is the
number of dimensions,
but it acts as the identity except on the $\mu$:th and $\nu$:th factors.
We can interpret $U_\mn(x)$ as the amplitude for parallel transport of
a string element across the plaquette.

\begin{figure}
\begin{center}
\includegraphics{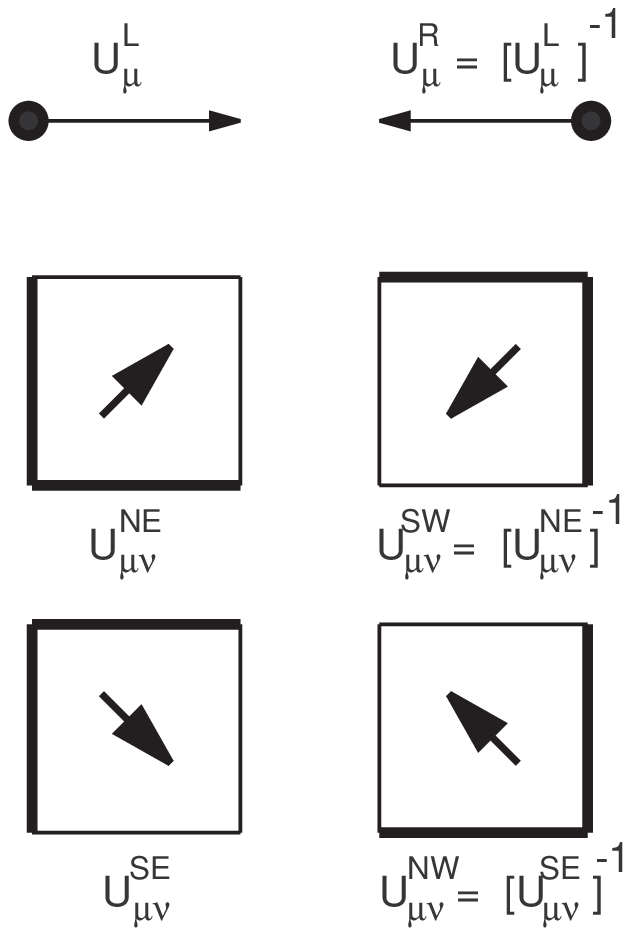}
  \caption{Upper row: $U_\mu^{L}(x)$ and 
  $U_\mu^{R}(x+\hmu) = [U_\mu^{L}(x)]^{-1}$.\hfill\break
  Middle row: $U_\mn^{NE}(x)$ and 
  $U_\mn^{SW}(x+\hmu+\hnu) = [U_\mn^{NE}(x)]^{-1}$.\hfill\break
  Lower row: $U_\mn^{SE}(x+\hnu)$ and 
  $U_\mn^{NW}(x+\hmu) = [U_\mn^{SE}(x+\hnu)]^{-1}$.}
    \label{fig:ampl}
  \end{center}
\end{figure}

In fact, and this is something which I missed in \cite{Lar90}, two 
independent variables are needed for each plaquette. 
$U_\mn(x) \equiv U_\mn^{NE}(x)$ can be
interpreted as the amplitude for parallel transport of a string element
across the NE diagonal, as illustrated in Figure \ref{fig:ampl}. 
It is then clear that four such amplitudes are needed, one each for the 
four different directed diagonals. However, only two amplitudes are 
independent, since $U_\mn^{SW}(x+\hmu+\hnu) = [U_\mn^{NE}(x)]^{-1}$
and $U_\mn^{SE}(x+\hnu) = [U_\mn^{NW}(x+\hmu)]^{-1}$.

\begin{figure}
\begin{center}
\includegraphics{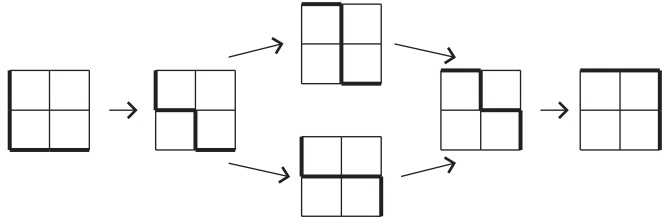}
  \caption{Associativity. The amplitude of parallel transport across a
  $2\times2$ square is independent of the intermediate steps.}
    \label{fig:assoc}
  \end{center}
\end{figure}

To every triangulated surface $\Si$, we can associate a holonomy 
$W(\Si)$ by contracting 
indices associated to links were two plaquettes are glued together. This
holonomy is the amplitude for transport of a string across the surface.
As we see in Figure \ref{fig:assoc}, there is a natural notion of
associativity. Since we now have a vector space $V_\ell$ associated to 
each link $\ell$ on the boundary of the Wilson surface, 
\be
W(\Si) \subset \bigotimes_{\ell\in\delta\Si} V_\ell.
\label{holonomy}
\ee

The curvature is the infinitesimal holonomy associated to each elementary 
cube. E.g., for a cube in the $123$-direction one has
\bes
&&UUUUUU = \\
&& U_{12}(x) U_{13}(x) U_{23}(x) [U_{12}(x+\three)]^{-1} 
[U_{13}(x+\two)]^{-1} [U_{23}(x+\one)]^{-1}
\eens
and the action reads
\be
\sum_{\hbox{elementary cubes}} \tr UUUUUU + \hbox{ $7$ more terms.}
\ee
The eight terms corresponds to the cube's eight directed diagonals.
Just as the two terms in ordinary lattice gauge theory can be
interpreted as parallel transport of a particle around the plaquette,
in the clockwise and counter-clockwise directions, the eight terms
here rotate a string piece around the cube's diagonals.

\begin{figure}
\begin{center}
\includegraphics{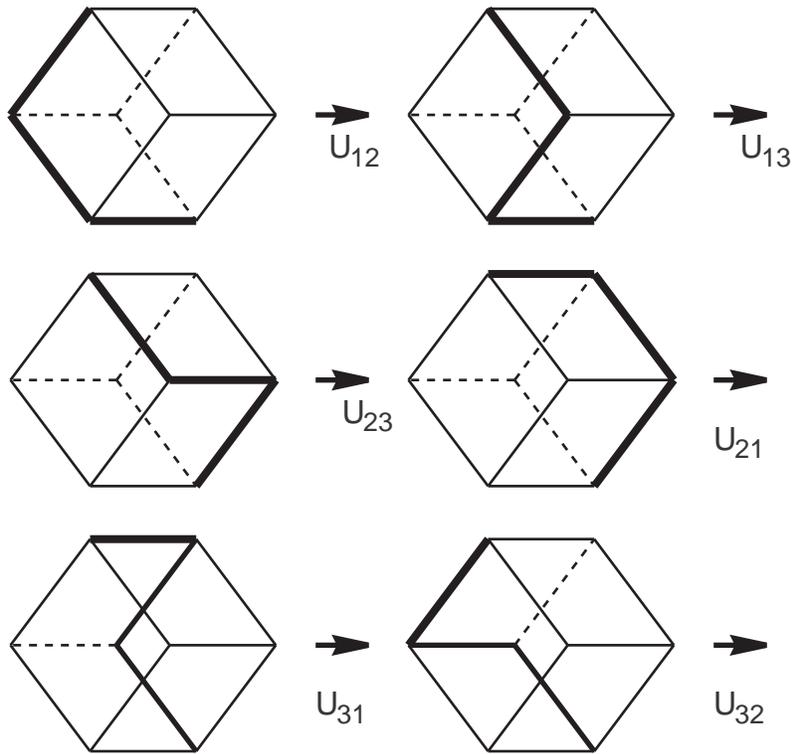}
  \caption{The curvature is the amplitude for parallel transport of string 
  element around a cube.}
    \label{fig:string}
  \end{center}
\end{figure}

There is a gauge symmetry associated to each link; the transformation
\be
U_\mn(x) \to [f_\mu(x+\hnu)]^{-1} [f_\nu(x+\hmu)]^{-1}
 U_\mn(x) f_\mu(x) f_\nu(x)
\ee
leaves the action invariant. The gauge invariant observables are 
closed Wilson surfaces, i.e. the product of four-index 
objects around a closed, two-dimensional surface. The natural continuum 
formulations of this model are in terms of loop or membrane variables 
(non-zero and zero curvature, respectively). Locality is not 
manifest in these formulations, but it is clear on the lattice that the 
model is perfectly local; the action is a sum over elementary cubes.

\begin{figure}
\begin{center}
\includegraphics{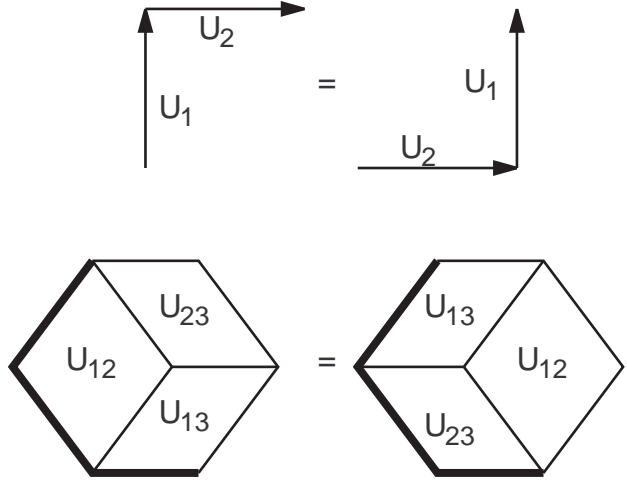}
  \caption{Commutativity $U_1 U_2 = U_2 U_1$ and
  the Yang-Baxter equation $U_{12} U_{13} U_{23} = U_{23} U_{13} U_{12}$.}
    \label{fig:yb}
  \end{center}
\end{figure}

The zero-curvature condition is quite interesting. In the spatially
homogeneous case (objects depend on orientation but not on location),
zero 1-curvature becomes $U_1 U_2 U_1^{-1} U_2^{-1} = 0$, i.e. $U_1$ and
$U_2$ commute. Vanishing $2$-curvature becomes
\be
U_{12} U_{13} U_{23} = U_{23} U_{13} U_{12},
\ee
which is the Yang-Baxter equation, of paramount importance to the
theory of integrable lattice models in two dimensions. The standard
illustration of the Yang-Baxter equation is the equality of two cube
halves, as shown in Figure \ref{fig:yb}.

It is immediate how to generalize this model to $p$-Yang Mills theory 
on the lattice. The zero $p$-curvature condition is known as the
$p$-simplex equation. It is sufficient to construct integrable lattice
models in $p$ dimensions, but unfortunately no interesting solutions
to it are known. More precisely, some solutions are known, but one needs a
continuum of solutions depending on variables like temperature or
magnetic field to be able to compute critical exponents.

\section{Gerbes}

\subsection{Bundles}

Let us first review the definition of an ordinary line bundle over an 
$n$-dimensional manifold $M$. Start with a good cover of $M$. Each 
neighborhood $U_a$ looks like $\RR^n$, and on the overlaps $U_a \cap U_b$
we define transition functions $g_{ab}$ with values in $S^1$. We can 
illustrate each neighborhood by a $\bullet$ and the overlap, or rather 
the transition function, as an arrow between bullets:
\[
\parbox{44mm}{\begin{picture}(84,15)
\put(1,10){$U_a$}
\put(2,1){$\bullet$}
\put(4,3){\vector(1,0){78}}
\put(84,1){$\bullet$}
\put(84,10){$U_b$}
\put(39,10){$g_{ab}$}
\end{picture}} 
\]
The transition functions must satisfy the consistency conditions
\bes
 1)&\qquad&  g_{ab} g_{ba} = 1, \nle
 2)&&  g_{ab} g_{bc} g_{ca} = 1,
\eens
which can be illustrated by the following diagrams
\[
\parbox{44mm}{\begin{picture}(85,35)
\put(1,25){$U_a$}
\put(2,11){$\bullet$}
\put(4,18){\vector(1,0){78}}
\put(85,10){\vector(-1,0){78}}
\put(83,11){$\bullet$}
\put(83,25){$U_b$}
\put(38,25){$g_{ab}$}
\put(38,0){$g_{ba}$}
\end{picture}} 
\]
\[
\parbox{44mm}{\begin{picture}(85,120)
\put(1,102){$U_a$}
\put(2,91){$\bullet$}
\put(4,93){\vector(1,0){78}}
\put(85,91){$\bullet$}
\put(85,102){$U_b$}
\put(39,102){$g_{ab}$}
\put(42,12){$\bullet$}
\put(42,0){$U_c$}
\put(87,93){\vector(-1,-2){38}}
\put(70,50){$g_{bc}$}
\put(44,15){\vector(-1,2){37}}
\put(3,50){$g_{cb}$}
\end{picture}} 
\]
i.e. going round a triangle results in the unit operator. These conditions 
make $g_{ab}$ into a Cech cocycle in $H^0(M, C^\infty(S^1))$.

Two manifolds $g_{ab}$ and $g'_{ab}$ are equivalent if there exist
functions $f_a$ on $U_a$ such that
\be
   g'_{ab} = g_{ab} f_a^{-1} f_b
\ee
corresponding to the picture
\[
\parbox{44mm}{\begin{picture}(84,15)
\put(1,10){$f_a^{-1}$}
\put(2,1){$\bullet$}
\put(4,3){\vector(1,0){78}}
\put(84,1){$\bullet$}
\put(84,10){$f_b$}
\put(39,10){$g_{ab}$}
\end{picture}} 
\]
This is already very reminiscent of a lattice gauge theory with 
neighborhoods playing the role of points and transition functions the
role of amplitudes, and the equivalence relation is recognized as a 
gauge transformation.

\subsection{Gerbes}

A gerbe (more precisely, 1-gerbe) is a generalizion of a bundle. On
every triple overlap $U_a \cap U_b \cap U_c$ we define a function 
$g_{abc}$, which can be illustrated by the oriented triangle
\[
\parbox{44mm}{\begin{picture}(85,120)
\put(1,102){$U_a$}
\put(2,91){$\bullet$}
\put(4,93){\vector(1,0){78}}
\put(85,91){$\bullet$}
\put(85,102){$U_b$}
\put(36,65){$g_{abc}$}
\put(42,12){$\bullet$}
\put(42,0){$U_c$}
\put(87,93){\vector(-1,-2){38}}
\put(44,15){\vector(-1,2){37}}
\end{picture}} 
\]
The transition functions satisfy
\be
  g_{abc} = g_{bca} = g_{bac}^{-1}
\ee
and the cocycle condition on quadruple overlaps
\be
  g_{abc} g_{abd}^{-1} g_{acd} g_{bcd}^{-1} = 1.
\ee
This condition corresponds to a tetrahedron diagram. Two gerbes $g_{abc}$
and $g'_{abc}$ are equivalent if there are functions $f_{ab}$ living on 
double overlaps (i.e. the edges of the triangle), such that
\be
  g'_{abc} = g_{abc} f_{ab} f_{bc} f_{ca}.
\label{gerbeeq}
\ee
These conditions make $g_{abc}$ into a Cech cocycle in 
$H^1(M, C^\infty(S^1))$.
Again there is striking resemblance with the lattice 2-gauge theory I
described: the transition function $g_{abc}$ lives on a plaquette, there 
is a gauge invariance $f_{ab}$ living on the link, and a curvature 
associated to a $3$-dimensional cell. The main difference is that two-form
lattice gauge theory involves square plaquettes whereas the gerbe picture 
gives rise to triangles.

It is now straightforward to extend the definitions to $p$-gerbes in terms
of functions living on $(p+2)$-fold overlaps and satisfying cocycle 
conditions on $(p+3)$-fold overlaps. In particular, bundles = 0-gerbes and
gerbes = 1-gerbes.

\subsection{Connections on abelian bundles}

Define a one-form $A_a$ on $U_a$ and a global two-form $F$ by
\be
\barr{lll}
  A_a - A_b = d \log g_{ab}    
  &\qquad&\hbox{on } U_a \cap U_b \\
  F = dA_a                  
  &\qquad&\hbox{on } U_a
\earr
\ee
Two bundles $g_{ab}$ and $g'_{ab}$ with connections $A_a$ and $A'_a$
are equivalent if $g_{ab} \sim g'_{ab}$ with equivalence $f_a$ 
(\ref{gerbeeq}) and
\be
  A'_a = A_a + d \log f_a    
  \qquad\hbox{on } U_a
\ee

\subsection{Connections on abelian gerbes}

Define a one-form $A_{ab} = -A_{ba}$ on $U_a \cap U_b$, a two-form 
$F_a$ on $U_a$ and a global three-form $G$.
\be
\barr{lll}
  A_{ab} + A_{bc} + A_{ca} = d \log g_{ab}
  &\qquad&\hbox{on } U_a \cap U_b \cap U_c \nl
  F_b - F_a = d A_{ab}                 
  &\qquad&\hbox{on } U_a \cap U_b \\
  G = dF_a                    
  &\qquad&\hbox{on } U_a    
\earr
\ee
Two gerbes $g_{abc}$ and $g'_{abc}$ with gerbe-connections $A_{ab}$, 
$F_a$ and $A'_{ab}$, $F'_a$ are equivalent if $g_{abc} \sim g'_{abc}$ and 
\be
\barr{lll}
  A'_{ab} = A_{ab} + B_b - B_a - d \log f_{ab}    
  &\qquad&\hbox{on } U_a \cap U_b \nle
  F'_a = F_a + dB_a           
  &\qquad&\hbox{on } U_a.
\earr
\ee
Note the trade-off between overlap and spacetime indices (form degree).

\subsection{Connection on non-abelian gerbes}

The treatment of abelian gerbes seems rather well established. However,
if one wants to identify the Yang-Baxter equation with a flatness 
condition for gerbe connection, one must consider non-abelian gerbes;
the Yang-Baxter equation $R_{12} R_{13} R_{23} = R_{23} R_{13} R_{12}$
is not very interesting if $R_{ab}$ is abelian. 

Mackaay \cite{Mac01} defines local one-forms $A_a \in U_a$, valued in 
the Lie algebra $\oj$ of $G$, such that
\be
  A_b - f_{ba} A_a f_{ab} - d \log f_{ab} = A_{ab} 
  \qquad\hbox{on } U_{ab}.
\ee
Two gerbes $f_{ab}$, $g_{abc}$ and $f'_{ab}$, $g'_{abc}$ with connections 
$A_a$, $A_{ab}$, $F_a$ and $A'_a$, $A'_{ab}$, $F'_a$ are equivalent if 
there exists $h_a$ such that
\be
  A'_a = h_a^{-1} A_a h_a + B_a + d \log h_a.
\ee

\subsection{Local expressions for gerbe connections and curvatures}

Locally, a bundle connection is a $\oj$-valued one-form $A = A_\mu dx^\mu$ 
and the curvature is a $\oj$-valued two-form $F = F_\mn dx^\mu dx^\nu$,
where $F = dA + [A,A]$. 
The Bianchi identity reads $dF + [A,F] = 0$.

Attal \cite{Att02} introduces the following data\footnote{Attal uses
different letters.} to describe a gerbe
connection: a $\oj$-valued one-form $A$ (connection) and an 
$\der(\oj)$-valued two-form $B$ (curving), whose curvatures are given by
a $\oj$-valued two-form $F$ and a $\der(\oj)$-valued three-form $G$,
where
\bes
F &=& dA + [A,A] + B, \nl
G &=& dB + [A,B], 
\label{Attal}\\
{[}B,B] &=& 0.
\eens
The Bianchi identity
\be
dG + [A,G] = [F,B]
\ee
follows from $d^2 = 0$ and the Jacobi identity.

However, this formulation seems unsuitable to describe the generalized
lattice model in Section 2, for several reasons.
\begin{itemize}
\item
A hierarchy of connections arises in the continuum description but only
the top $(p+1)$-form connection appears on the lattice. 
\item
The modified curvature $F' = F-B = dA + [A,A]$ is the usual curvature
of the one-form connection $A$.
\item
There is no natural covariant derivative.
\item
There seems to be no natural place for the Yang-Baxter equation.
\end{itemize}
With these problems in mind, I propose an alternative local formulation
of gerbes in the next section.

\section{Gerbes in local coordinates}

\subsection{Symmetries}

Locally, the gerbe coordinates are $(x,s)$, where $x = (x^\mu)$, 
$\mu = 1, 2, ..., n$ is a point in the underlying space $\RR^n$, and
$s = (s^{\mu_1\mu_2..\mu_p})$, is a $p$-dimensional surface element.
We have $s^{\mu_1..\mu_i..\mu_j..\mu_p} = -s^{\mu_1..\mu_j..\mu_i..\mu_p}$.
For definiteness, we will mainly consider the case $p=2$ in the sequel,
but all formulas are readily generalized to arbitrary $p$, and later 
the case $p=1$ will receive special attention.
Introduce derivatives $\dmu = \d/\d x^\mu$ and $\ethmn = \d/\d s^\mn$,
satisfying the Heisenberg algebra
\be
[\dnu, x^\mu] = \dlt^\mu_\nu, \qquad
[\ethrs, s^\mn] = \dlt^\mu_\rho \dlt^\nu_\si 
- \dlt^\mu_\si \dlt^\nu_\rho.
\ee

Let $\xi = \xmu(x)\dmu$ be a vector field acting on the base space 
$\RR^n$. The algebra of vector fields $\vect(n)$, i.e. the algebra of 
infinitesimal diffeomorphisms, can be realized as 
\bes
\Lxi &=&
\xmu(x)\dmu + \dnu\xmu(x)s^\nr\ethmr + \dnu\xmu(x)T^\nu_\mu 
\nlb{Lxi}
&=& \xmu(x) + \dnu\xmu(x)\tT^\nu_\mu, 
\eens
where both
\be
\tT^\mu_\nu \equiv s^\mr\ethnr + T^\mu_\nu
\label{tT}
\ee
and $T^\mu_nu$ satisfy $gl(n)$:
\be
[\tT^\mu_\nu, \tT^\rho_\si] = \dlt^\rho_\nu \tT^\mu_\si 
 - \dlt^\mu_\si \tT^\rho_\nu.
\label{gln}
\ee
Let $J^a$ be the generators of a finite-dimensional Lie algebra $\oj$
with structure constants $f^{ab}{}_c$, i.e.
\be
[J^a, J^b] = f^{ab}{}_c J^c.
\label{oj}
\ee
Then $X = X_a(x,s)J^a$ generates an algebra of gerbe gauge transformations, 
which we call the {\em gerbe gauge algebra} and denote by 
$\mapp$. Note that $X$ depends on both the base point $x$ and the surface 
element $s$, but $\xi$ only depends on $x$.
The brackets in $\vmap$ read
\bes
[\xi, \eta] &=& \xmu(x)\dmu\ynu(x)\dnu - \ynu(x)\dnu\xmu(x)\dmu, \nl
{[}\xi, X] &=& (\xmu(x)\dmu X_a(x,s) 
+ \dnu\xmu(x)s^\nr\ethmr X_a(x,s))J^a, 
\label{xX}\\
{[}X,Y] &=& f^{ab}{}_c X_a(x,s) Y_b(x,s) J^c.
\eens
Observe the second term in the middle equation, which is absent for
ordinary gauge transformations. 
The Lie derivatives (\ref{Lxi}) and $\JX = X_a(x,s)J^a$ satisfy the
same algebra:
\bes
[\Lxi, \L_\eta] &=& \L_{[\xi,\eta]}, \nl
{[}\Lxi, \JX] &=& \J_{[\xi,X]}, 
\label{LJ}\\
{[}\JX, \J_Y] &=& \J_{[X,Y]},
\eens
provided that $[T^\mu_\nu, J^a] = 0$.

\subsection{Sections and connections}

A gerbe section corresponds locally to a tensor field $\phi(x,s)$ valued
in a $\oj$ module. If the $gl(n)$ action is given by $T^\mu_\nu$ and the 
$\oj$ action by $J^a$, then $\phi$ carries the following representation
of $\vmap$:
\bes
\Lxi\phi &=& -\xmu\dmu\phi - \dnu\xmu s^\nr\ethmr\phi 
 - \dnu\xmu T^\nu_\mu \phi, 
\nlb{LJf}
\JX\phi &=& -X_aJ^a\phi.
\eens
Here and henceforth we suppress arguments, and keep in mind that all
fields and functions ($\phi$, $X_a$, etc) depend on $(x,s)$, except 
vector fields $\xmu$ which only depend on $x$.
It follows immediately from (\ref{LJf}) that the derivative $\dnu\phi$
transforms as
\bes
\Lxi\dnu\phi &=& -\xmu\dmu\dnu\phi - \dsi\xmu s^{\si\rho}\ethmr\dnu\phi 
 - \dsi\xmu T^\si_\mu \dnu\phi \nl
 && -\dnu\xmu\dmu\phi 
 - \dnu\dsi\xmu s^{\si\rho}\ethmr\phi - \dnu\dsi\xmu T^\si_\mu \phi,
\nlb{LJdf}
\JX\dnu\phi &=& -X_aJ^a\dnu\phi - \dnu X_aJ^a\phi.
\eens
Now define the covariant derivative
\be
D_\nu\phi = \dnu\phi + A_{a\nu}J^a\phi,
\label{Dnu}
\ee
which is covariant only w.r.t. $\mapp$, and
\be
\nabla_\nu\phi = \dnu\phi 
 + \Gamma^\si_{\tau\nu}\tT^\tau_\si\phi + A_{a\nu}J^a\phi, 
\label{Nnu}
\ee
which is covariant w.r.t. all of $\vmap$. The connections 
$A_{a\nu}(x,s)$ and $\Gamma^\si_{\tau\nu}(x,s)$ (which both depend on
both $x$ and $s$) transform as
\bes
\JX A_{a\nu} &=& -f^{bc}{}_a X_bA_{c\nu} + \dnu X_a, 
\nlb{JLA}
\Lxi A_{a\nu} &=& -\xmu\dmu A_{a\nu} 
 - \dsi\xmu s^{\si\rho}\ethmr A_{a\nu}- \dnu\xmu A_{a\mu}, 
\nnl
\JX \Gamma^\si_{\tau\nu} &=& 0, \\
\Lxi \Gamma^\si_{\tau\nu} &=& -\xmu\dmu\Gamma^\si_{\tau\nu}
 - \d_\ka\xmu s^{\ka\rho}\ethmr\Gamma^\si_{\tau\nu}
 - \d_\ka\xmu (T^\ka_\mu \Gamma)^\si_{\tau\nu}
 + \dtau\dnu\xi^\si.
\eens
The consistency of these transformation laws follow immediately because
both the covariant derivative and the ordinary derivative transform
consistently, and hence so does their difference.

For brevity, we sometimes write $A_\nu =  A_{a\nu}J^a$,
$\Gamma_\nu = \Gamma^\si_{\tau\nu}\tT^\tau_\si$,  and
$\nabla_\nu = \dnu + \Gamma_\nu + A_\nu$. The associated curvature is
\be
[\nabla_\mu, \nabla_\nu] = R_\mn  + F_\mn,
\ee
where
\bes
R_\mn  &=& \dmu\Gamma_\nu - \dnu\Gamma_\mu 
 + [\Gamma_\mu,\Gamma_\nu], \nl
F_\mn  &=& \dmu A_\nu - \dnu A_\mu + [\Gamma_\mu, A_\nu]
  - [\Gamma_\nu, A_\mu] + [A_\mu, A_\nu] 
\label{F}\\
&\equiv& \nabla_\mu A_\nu - \nabla_\nu A_\mu + [A_\mu, A_\nu].
\eens
Note that the derivative of $A_\nu(x,s)$ need covariantization because 
it is a ($\oj$-valued) function of both $x$ and $s$:
\be
\nabla_\mu A_\nu = \dmu A_\nu 
 + \Gamma^\si_{\tau\mu}s^{\tau\rho}\eth_{\si\rho} A_\nu.
\ee

Apart from the ordinary derivative w.r.t. $x$, we must also convariantize
the gerbe derivative $\ethst$. {F}rom
\bes
\Lxi\ethst\phi &=& -\xmu\dmu\ethst\phi 
 - \dnu\xmu s^\nr\ethmr\ethst\phi - \dnu\xmu T^\nu_\mu\ethst\phi \nl
 &&- \dsi\xmu \eth_\mt\phi - \dtau\xmu \eth_\sm\phi, \\
\JX\phi &=& -X_aJ^a\ethst\phi - \ethst X_aJ^a\phi,
\eens
we see that $\ethst\phi$ transforms under $\vect(n)$ as a tensor with 
two extra indices, but the $\mapp$ action needs compensation.
Introduce
\be
\Delta_\st\phi  = \ethst\phi + B_{a\st}(x,s)J^a\phi
 \equiv \ethst\phi + B_\st \phi.
\ee
It follows that
\bes
\Lxi B_{a\st} &=& -\xmu\dmu B_{a\st}
 - \dnu\xmu s^\nr\ethmr B_{a\st}
 - \dsi\xmu B_{a\mt} - \dtau\xmu B_{a\sm}, \nl
\JX B_{a\st} &=& -f^{bc}{}_a X_b B_{c\st} + \ethst X_a.
\ees
The associated curvature is
\be
[\Delta_\mn, \Delta_\st] = \ethmn B_\st  
 - \ethst B_\mn  + [B_\mn, B_\st].
\ee
One can also define the ``cross curvature'' 
$[\nabla_\mu, \Delta_\st]$.

\subsection{Rescaling invariance}

We can use the covariant gerbe derivative $\Delta_\st $ to impose
invariance under surface rescalings. One immediately checks that
$s^\st \Delta_\st \phi$ and $\phi$ transform in the same way
under $\vmap$. Therefore we can consistently impose the constraint
\be
s^\st \Delta_\st \phi = \la \phi
\label{scalefix}
\ee
for any constant $\la$.

Alternatively, we may consider invariance under rescalings of the 
surface elements as a symmetry. Consider gerbe sections $\phi(x,s)$ 
which are invariant under transformations of the form
$s^\mn \to f s^\mn$. Moreover, we let rescalings depend on the
base point $x$, so infinitesimally $s^\mn \to (1+F(x))s^\mn$ for some
function $F(x)$. The {\em rescaling} $F = F(x)$ satisfies
\bes
{[}\xi, F] &=& \xmu(x)\dmu F(x), \nl
{[}F,G] &=& 0, \\
{[}F, X] &=& s^\mn F(x) \ethmn X_a(x,s) J^a,
\eens
in addition to the brackets in (\ref{xX}). Eq. (\ref{LJ}) is 
supplemented by
\bes
[\Lxi, \RF] &=& \R_{[\xi,F]}, \nl
{[}\RF, \R_G] &=& 0, \\
{[}\RF, \JX] &=& \J_{[F,X]}.
\eens
In the parlance of constrained systems, $\RF$ generates a gauge
symmetry. The constraint $\RF = 0$ is first class, which together with
the gauge condition (\ref{scalefix}) becomes a second class constraint.

Introduce two generators $\La$ and $U^\rho_\st$, satisfying the algebra
\bes
[\La, U^\rho_\st] &=& -2U^\rho_\st, \nl
{[}\La, T^\rho_\si] &=& {[}\La, \La] = [U^\ka_\mn, U^\rho_\st] = 0, \\
{[}T^\mu_\nu, U^\rho_\st] &=& \dlt^\rho_\nu U^\mu_\st
 - \dlt^\mu_\si U^\rho_{\nu\tau} - \dlt^\mu_\tau U^\rho_{\si\nu}.
\eens
Then it turns out that the rescalings can be realized as
\be
\RF = F(x)s^\mn\ethmn + F(x)\La + \drho F(x) s^\st U^\rho_\st.
\label{RF}
\ee
To prove that (\ref{RF}) indeed furnishes a realization of the
rescaling algebra, it is useful to introduce the abbreviations
\be
\tLa = s^\mn\ethmn + \La, \qquad
\tU^\rho = s^\st U^\rho_\st.
\ee
It is clear that
\bes
[\tLa, \tU] &=& [\tLa, \tT^\rho_\si] = {[}\tLa, \tLa] 
= [\tU^\mu, U^\nu] = 0, \nle
{[}\tT^\mu_\nu, \tU^\rho] &=& \dlt^\rho_\nu \tU^\mu, 
\eens
and that we can rewrite (\ref{RF}) as
\be
\RF = F(x) \tLa + \drho F(x) \tU^\rho.
\ee

\subsection{A special gauge choice}

Assume that we require
\be
A_{a\mu}(x,s) = A_{a\mnr}(x) s^\nr,
\label{As}
\ee
where $A_{a\mnr} = A_{a\nr\mu} = -A_{a\nu\mr}$. In 
coordinate-free notation \break
$A = A_{a\mnr} dx^\mu dx^\nu dx^\rho J^a$ 
is a $\oj$-valued three-form. This choice is obviously not preserved by
$\mapp$, but we may hope that it preserved by the subalgebra
generated by $X$'s of the form
\be
X_a(x,s) = X_{a\mn}(x)s^\mn.
\label{Xs}
\ee
The transformation law (\ref{JLA}) becomes
\be
s^\nr\JX A_{a\mnr} = 
  -f^{bc}{}_a X_{b\st} s^\st  s^\nr A_{c\mnr}
 + s^\nr \d_{[\mu}X_{|a|\nr]},
\ee
where
\be
\d_{[\mu}X_{|a|\nu\rho]} = 
 2( \dmu X_{a\nu\rho} +  \dnu X_{a\rho\mu} +  \drho X_{a\mu\nu} ).
\ee
A sufficient condition for this to hold is clearly 
\be
\JX A_{a\mnr} = 
  -f^{bc}{}_a X_{b\st} s^\st  A_{c\mnr}
 + \d_{[\mu}X_{|a|\nu\rho]}.
\label{JA3}
\ee
In coordinate-free notation, this reads $\JX A = [s\cdot X,A] + dX$,
where $s\cdot X$ is the contraction of the two-form $X$ and the 
two-vector $s$.
The associated curvature can now be written
\be
F_\mn (x,s) = F_{\mu\nu\rs}(x) s^\rs,
\label{Fs}
\ee
where $F_{\mu\nu\rs}$ is totally anti-symmetric in spacetime
indices and
\be
F_{\mu\nu\rs} = \d_{[\mu}A_{\nu\rs]} 
+ [A_{[\mnr}, s^{\ka\tau} A_{\si]\ka\tau}].
\label{F4}
\ee
In coordinate-free notation, the four-form $F = dA + [A, s\cdot A]$
transforms as $\JX F = [X,F]$ under $\mapp$.

However, the consistency of (\ref{As}) requires not only (\ref{JA3}),
but it also relies on the assumption that the gauge choice (\ref{Xs}) 
defines a subalgebra. One checks that the commutator of two fields of
the form (\ref{Xs}),
\be
[X,Y] = s^\mn s^\rs X_{a\mu\nu}(x) Y_{b\rs}(x) f^{ab}{}_c J^c,
\ee
is in general {\em not} of the same form unless $f^{ab}{}_c = 0$. 
Hence the choice (\ref{As}) is not invariant even under 
the subalgebra (\ref{Xs}) of $\mapp$ if $\oj$ is non-abelian.

\subsection{ Gerbe holonomy}
Let $\Si$ be a closed three-dimensional manifold with local coordinates 
$(\si,\tau)$, where $\si = (\si^1, \si^2)$. We have thus choosen a 
foliation of $\Si$. Now regard $\Si$ as a submanifold 
$\Si\subset\RR^n$, with the
embedding given by coordinate functions $x^\mu = x^\mu(\si,\tau)$. 
The surface element becomes on $\Si$
\be
s^\mn \equiv s^\mn(\si,\tau) = 
\eps^{ij}{\d x^\mu\/\d\si^i}{\d x^\nu\/\d\si^j}.
\ee
If $\oj$ is abelian, we define the holonomy 
\be
W(\Si) = \exp(\int_\Si \dst A_\mu(x,s) \dxmdt).
\ee
One checks that $W(\Si)$ is invariant under the gerbe gauge algebra:
\bes
\JX W &=& W(\Si) \int_\Si \dst \dmu X(x,s)  \dxmdt \nle
&=& W(\Si) \int_\Si \dst {\d\/d\tau}X(\si,\tau) = 0,
\eens
since the $\tau$ integral vanishes. Diffeomorphism invariance is also
clear.

If we make the special gauge choice (\ref{As}), the holonomy becomes
\bes
W(\Si) &=& \exp(\int_\Si \dst A_\mnr(x){\d x^\mu\/\d\tau}
\eps^{ij} {\d x^\nu\/\d\si^i} {\d x^\rho\/\d\si^j} ) \nl
&=& \exp(c \int_\Si d^3\!x\ A_\mnr(x)\eps^\mnr) 
\label{W3}\\
&=& \exp(c \int_\Si A).
\eens
since
\be
{\d x^{[\mu}\/\d\tau}{\d x^\nu\/\d\si^1} {\d x^{\rho]}\/\d\si^2}
= c \eps^\mnr \times \hbox{Jacobian}
\ee
for some constant $c$.
The holonomy (\ref{W3}) is invariant under the restricted set of gauge 
transformations (\ref{Xs}), because
\be
\JX W(\Si) \propto W(\Si) \int_\Si dX \propto 
W(\Si) \int_{\delta\Si} X,
\ee
and the boundary $\delta\Si = 0$ by assumption.

Let us now turn to the case $\oj$ non-abelian. In the ordinary one-form 
gauge theory case, the exponential of the integral must be replaced by 
the path-ordered integral:
\be
W(C) = \P \exp(\int_C d\tau\ A_\mu(x(\tau)) \dxmdt(\tau)),
\ee
where $C$ is some curve. This formal expression is most
intuitively defined in the lattice approximation. In particular, $W(C)$
satisfies the relation $W(C_1)W(C_2) = W(C_1\circ C_2)$, where
$C_1\circ C_2$ is the concatenation of $C_1$ and $C_2$. 

Analogously, we now define the surface-ordered integral
\be
W(\Si) = \P \exp(\int_{\dlt\Si} \dst A_\mu(x,s) \dxmdt)
\label{WSi}
\ee
by its lattice regularization. In particular, $W(\Si)$ takes values in 
the space $V^{\otimes\int_{\dlt\Si} d^2\!\si}$, which should be thought
of as the continuum analogue of (\ref{holonomy}). The size of the space 
(a continuum tensor product) makes the expression (\ref{WSi}) merely
formal, in contrast to the manifestly well-defined local expressions like
(\ref{Dnu}) and (\ref{F}).

\subsection{ Gerbe Yang-Mills theory }

In this subsection we assume that there is a preserved constant (and 
thus flat) metric $g_\mn $ with inverse $g^\mn $, and hence that 
diffeomorphism invariance is broken down to Poincar\'e invariance. 
This means that $\Gamma^\si_{\tau\nu} = 0$, so we can ignore the
difference between the covariant derivatives $D_\nu$ and $\nabla_\nu$
(\ref{Dnu}), (\ref{Nnu}).

The natural gerbe generalization of the pure Yang-Mills action is
\be
S = {1\/2} \int d^n\!x\ \int d^2\!s\ F^{a\mu\nu}(x,s) F_{a\mu\nu}(x,s),
\ee
which leads to the Yang-Mills equations
\be
D_{a\nu} F^{a\mu\nu}(x,s) = 0.
\label{YM}
\ee
In particular, if we assume that the connection is of the form (\ref{As}),
we recover the equations of motion of $(p+1)$-form electromagnetism
in the abelian case:
\be
s_\rs \dnu F^{\mu\nu\rs}(x) = 0.
\ee
The gerbe equations (\ref{YM}) are thus the natural non-abelian 
generalization of $(p+1)$-form electromagnetism.

\subsection{ The classical Yang-Baxter equation }

In the previous subsections all formulas were specialized to the case
$p=2$, but the analogous formulas for arbitrary $p$ are readily deduced.
We here set $p=1$, so the line coordinate $s = (s^\mu)$ becomes a
one-vector. If we make the gauge choice analogous to (\ref{As}), so the
connection is given by a two-form $A_\mn (x)$, the curvature 
three-form becomes
\be
F_\mnr(x) = \d_{[\mu}A_{\nr]}(x) 
+ [A_{[\mn}(x), s^{\si} A_{\rho]\si}(x)].
\label{F3}
\ee
This gauge choice is of course completely non-invariant.

Let $A_\mn (x)$ be spatially homogeneous, i.e. it does not depend
on $x$ at all, so the first term above vanishes. Moreover, consider
the special point $s = (1,1,1)$ in three dimensions. The zero-curvature 
condition $F_\mnr = 0$ then exlicitly becomes
\bes
[A_{12}, A_{31} + A_{32}] +
[A_{23}, A_{12} + A_{13}] +
[A_{31}, A_{21} + A_{23}] &=& \nle
-2( [A_{12}, A_{13}] + [A_{12}, A_{23}] + [A_{13}, A_{23}] ) &=& 0.
\eens
This is recognized as the classical Yang-Baxter equation (CYBE).
That this equation arises in the continuum formulation is hardly 
surprising since we have seen that the QYBE appears on the lattice.

As is well known, the CYBE is an equation on the triple space
$V^{\otimes3} = V_1\otimes V_2\otimes V_3$, where $V$ is a vector space
associated to a link, and $A_\mn $ acts non-trivially on
$V_\mu\otimes V_\nu$ only: $A_{12} = A\otimes\id$, etc.
The simplest class of solutions are the trigonometric ones of the form
\be
A(u,v) = {1\/u-v} J^a\otimes J_a,
\ee
where $J^a$ are the generators of $\oj$ and we have contracted indices 
using the Killing metric on $\oj$. Note that the solution depends
on additional, ``spectral'' parameters. If we collect these into a vector
$u = (u_1, u_2, u_3)$ we may write
\be
A_{12}(u) = {1\/u_1-u_2} J^a\otimes J_a\otimes\id,
\ee
etc.

Returning to $p=2$, vanishing of the curvature four-form (\ref{F4}),
$F_{\mu\nu\rs} = 0$, becomes in the spatially homogeneous case
and at the point 
$s^{12} = s^{13} = s^{14} = s^{23} = s^{24} = s^{34} = 1$:
\bes
[A_{123}, A_{124}] + [A_{123}, A_{134}] + [A_{123}, A_{234}] &+& 
\nlb{CTetra}
[A_{124}, A_{134}] + [A_{124}, A_{234}] + [A_{134}, A_{234}] &=& 0.
\eens
This is the infinitesimal form of Zamolodchikov's tetrahedron equation
\cite{BS84,Bax83,BF86,Zam81}. It is an equation in the space 
$V^{\otimes6} =  V_{12}\otimes V_{13}\otimes V_{14}\otimes V_{23}
\otimes V_{24}\otimes V_{34}$, such that $A_\mnr$ acts trivially 
on all spaces except $V_\mn \otimes V_\mr\otimes V_\nr$.

The classical tetrahedron equation (\ref{CTetra}) has a rather disquiting
property. Of the six terms, the only one that is non-trivial on all
factors except $V_{34}$ is $[A_{123}, A_{124}]$. Therefore, this 
expression can in fact not act non-trivially on all of
$V_{12}\otimes V_{13}\otimes V_{14}\otimes V_{23}\otimes V_{24}$,
but it rather consists of terms that act on four spaces only. The
natural ansatz is $A_{123} = B_{12} + B_{13} + B_{23}$, where 
$B_\mn $ is a solution of the Yang-Baxter equation. I am not aware of any
genuine solutions to (\ref{CTetra}).

The generalization to higher order is obvious.

\subsection{ Abelian extensions }

Upon quantization, we expect the gerbe gauge algebra to acquire an
abelian extension. To construct representations of such extensions
following \cite{Lar98}, we introduce a one-dimensional closed curve
$x^\mu = q^\mu(t)$, $s^\mn = \vsi^\mn(t)$, where $t\in S^1$, and expand 
all fields in a Taylor around $(x,s) = (q(t),\vsi(t))$:
\be
\phi(x,s) = 
\sum_{\scriptstyle|\mm|,|\rr|\atop \scriptstyle|\mm+\rr|\leq p}
{1\/\mm!\rr!} \phi_{\mm\rr}(t) (x-q(t))^\mm (s-\vsi(t))^\rr,
\ee
where $\mm = (m_1, .., \ab m_n)$ and $\rr = (r_{11}, .., r_{nn})$,
all $m_\mu\geq0$, $r_\mn\geq0$ ($\nu<\mu$) are multi-indices of length
$|\mm| = \sum_{\mu=1}^n m_\mu$ and 
$|\rr| = \sum_{\mu=1}^n \sum_{\nu=1}^\mu r_\mn$, respectively. Moreover,
\be
(x-q(t))^\mm = (x^1-q^1(t))^{m_1} ... (x^n-q^n(t))^{m_n},
\ee
and similar for $(s-\vsi(t))^\rr$.

After introducing canonical momenta for $q^\mu(t)$, $\vsi(t)$ and 
$\phi_{\mm\rr}(t)$ and normal ordering, we obtain Fock representations
of the following algebra:
\bes
[\Lxi, \L_\eta] &=& \L_{[\xi,\eta]} + {1\/2\pi i}\int dt
\ (c_1\dnu\dot\xmu(t)\dmu\ynu(t) +  c_2\dmu\dot\xmu(t)\dnu\ynu(t)), \nl
{[}\Lxi, \JX] &=& \J_{[\xi,X]}, 
\label{LJq}\\
{[}\JX, \J_Y] &=& \J_{[X,Y]} +
{k\/2\pi i}\int dt\ \dot X_a(t) Y_b(t) \dlt^{ab},
\eens
where $\dlt^{ab}$ is the Killing metric in $\oj$ and
\bes
\dnu\dot\xmu(t) &\equiv& \dot q^\rho(t)\dnu\drho\xmu(q(t)), \nle
\dot X_a(t) &\equiv& \dot q^\mu(t)\dmu X_a(q(t),\vsi(t)) + 
 \dot\vsi^\mn(t)\eth_\mn X_a(q(t),\vsi(t)).
\eens
Equation (\ref{LJq}) is the gerbe analogue of the Virasoro and affine
Kac-Moody algebras, and the abelian charges $c_1$, $c_2$ and $k$ can be
computed with the methods of \cite{Lar98}.

\section{Conclusion}
In this paper I have developed a local continuum formulation of the 
$p$-form lattice gauge theory in Section 2. Holonomy is an integral over
$p$-dimensional Wilson submanifolds, as in (\ref{holonomy}) and 
(\ref{WSi}), making this theory closely related to $(p-1)$-gerbes.
However, there are some advantages compared to other formulations:
\begin{itemize}
\item
A natural covariant derivative exists.
\item
There is only a $p$-form connection, not a hierarchy of connections.
\item
The relation to integrability, i.e. the quantum and classical Yang-Baxter
equations, is clear.
\end{itemize}
The main advantage compared to formulations in loop space is manifest
locality, which should facilitate explicit calculations.


\begin{thebibliography}{99}

\bibitem{AFSG97} O. Alvarez, L. A. Ferreira and J. S\`anchez Guill\'en,
  {\it A new approach to integrable theories in any dimension},
  {\tt hep-th/9710147} (1997).  

\bibitem{Att01} R. Attal,
  {\it Two-dimensional parallel transport: combinatorics and
  functoriality},
  {\tt math-ph/010505} (2001).

\bibitem{Att02} R. Attal,
  {\it Combinatorics of non-abelian gerbes with connection and
  curvature},
  {\tt math-ph/0203056} (2002).

\bibitem{BS84} V. V. Bazhanov and Yu. G. Stroganov,
  Nucl. Phys. {\bf B230 [FS10]} (1984) 435.

\bibitem{BD95} J. C. Baez and J. Dolan,
  {\it Higher-dimensional algebra and topological quantum field theory},
  J. Math. Phys {\bf36} (1995) 11.

\bibitem{Bax82} R. J. Baxter,
  {\it Exactly solved models in statistical mechanics},
  Academic Press, London (1982).

\bibitem{BF86} R. J. Baxter and P. J. Forrester,
  {\it Is the Zamolodchikov model critical?},
  J. Phys. A {\bf 18} (1986) 1483--1497.

\bibitem{Bax83} R. J. Baxter,
{\it On Zamolodchikov's solution of the tetrahedron equations},
  Comm. Math. Phys. {\bf 88} (1983) 185--205.

\bibitem{BM01} L. Breen and W. Messing,
  {\it Differential geometry of gerbes},
  {\tt math.AG/0106083} (2001).

\bibitem{Bry93} J.-L. Brylinski,
  {\it Loop spaces, characteristic classes and geometric 
  quantization},
  Prog. in Math. vol. 107, Birkh\"auser, Boston (1993).

\bibitem{CMR02} M. I. Caicedo, I. Martin and A. Restuccia,
  {\it Gerbes and duality},
  {\tt hep-th/0205002} (2002).

\bibitem{CMM97} A. L. Carey, J. Mickelsson and M. K. Murray,
  {\it Bundle gerbes applied to quantum field theory},
  {\tt hep-th/9711133} (1997).

\bibitem{Fre94} D. Freed,
  {\it Higher algebraic structures and quantization},
  Comm. Math. Phys. {\bf159} (1994) 343--398.

\bibitem{FN82} P. G. O. Freund and R. Nepomechie,
  Nucl. Phys. {\bf B199} (1982) 482.

\bibitem{Hit99} N. Hitchin,
  {\it Lectures on special Lagrangian submanifolds},
  {\tt math.DG/9907034} (1999).

\bibitem{Kog79} J. Kogut,
{\it An introduction to lattice gauge theory and spin systems},
  Rev. Mod. Phys. {\bf 51} (1979) 659--713.

\bibitem{Lar98} T. A. Larsson,
  {\it Extended diffeomorphism algebras and trajectories in jet space}.
  Commun. Math. Phys. {\bf 214} (2000) 469--491.

\bibitem{Lar90} T. A. Larsson,
  {\it $p$-cell gauge theories, manifold space and multi-dimensional 
  integrability},
  Mod Phys Lett A {\bf 5} (1990) 255--264.

\bibitem{MacPi01} M. Mackaay and R. Picken,
  {\it Holonomy and parallel transport for abelian gerbes},
  {\tt math.DG/0007053} (2001).

\bibitem{Mac01} M. Mackaay, 
  {\it A note on the holonomy of connetions in twisted bundles},
  {\tt math.DG/0106019} (2001).

\bibitem{Nep83} R. Nepomechie, Nuclear Physics B212 (1983) 310.

\bibitem{Orl83} P. Orland, Physics Letters 122B (1983) 78.

\bibitem{Orl84a} P. Orland in 
Gauge Theory on Lattice: 1984, Proceedings of the
Argonne Ntional Laboratory Workshop, National Technical Information
Service, Springfield, VA, USA (1984), page 305.

\bibitem{Orl84b} P. Orland, Imperial College preprint, July 1984. 
{\tt http://ccdb3fs.kek.jp/cgi-bin/img\_index?8408054}

\bibitem{Pol79} A. M. Polyakov,
  Phys. Lett. {\bf 82B} (1979) 247;
  Nucl. Phys. {\bf B164} (1979) 171.

\bibitem{Pol87} A. M. Polyakov,
  {\it Gauge fields and strings},
  Harwood, Chur (1987).

\bibitem{Tei86} C. Teitelboim,
  Phys. Lett. {\bf 167B} (1986) 63.

\bibitem{Zam81} A. B. Zamolodchikov,
  {\it Tetrahedron equations and the relativistic S-matrix of 
  straight-strings in 2+1-dimensions},
  Comm. Math. Phys. {\bf 79} (1981) 489 -- 505.

\bibitem{Zun00} Y. Zunger,
  {\it p-Gerbes and Extended Objects in String Theory},
  {\tt hep-th/0002074} (2000)


\end{thebibliography}
\end{document}